\documentstyle[twoside]{article}
\newcommand{\ie}{{\em i.e.}}
\newcommand{\eg}{{\em e.g.}}
\newcommand{\cf}{{\em cf. }}
\newcommand{\rhs}{{\em rhs }}
\newcommand{\re}{{\rm Re\,}}
\newcommand{\R}{I\!\!R}
\newcommand{\BB}{{\cal B}}
\newcommand{\OO}{{\cal O}}

\begin{document}

\title{Anomalous electron trapping by magnetic flux tubes and
electric current vortices}
\author{F. Bentosela,$^{a,b}$ P. Exner,$^{c,d}$ and
V.A. Zagrebnov$^{a,b}$}
\date{}
\maketitle
\begin{quote}
{\small \em a) Centre de Physique Th\'{e}orique, C.N.R.S.,
F--13288 Marseille Luminy \\
b) Universit\'{e} de la Mediterran\'{e}e (Aix--Marseille II),
F--13288 Luminy
\\ c) Nuclear Physics Institute,
Academy of Sciences, CZ--25068 \v Re\v z \\ d) Doppler Institute,
Czech Technical Univ., B\v rehov{\'a} 7, CZ-11519 Prague \\ \rm
\phantom{e)x}bentosela@cpt.univ-mrs.fr, exner@ujf.cas.cz,
zagrebnov@cpt.univ-mrs.fr} \vspace{8mm}

\noindent {\small We consider an electron with an anomalous
magnetic moment, $\,g>2\,$, confined to a plane and interacting
with a nonhomogeneous magnetic field $\,B\,$, and investigate the
corresponding Pauli Hamiltonian. We prove a lower bound on the
number of bound states for the case when $\,B\,$ is of a compact
support and the related flux is
$\,N+\epsilon\,,\;\epsilon\in(0,1]\,$. In particular, there are at
least $\,N+1\,$ bound states if $\,B\,$ does not change sign. We
also consider the situation where the magnetic field is due to a
localized rotationally symmetric electric current vortex in the
plane. In this case the flux is zero; there is a pair of bound
states for a weak coupling, and higher orbital-momentum
``spin-down" states appearing as the current strength increases.}
\end{quote}

\vspace{3mm}

\noindent Interaction of electrons with a localized magnetic field
has been a subject of interest for a long time. It has been
observed recently that a magnetic flux tube can bind particles
with spin antiparallel to the field provided the latter have an
anomalous magnetic moment, $\,g>2\,$. Recall that this is the case
for a free electron which has $\,g=2.0023$. The effect was
demonstrated first in simple examples \cite{CFC,Mo}, notably those
of a circular tube with a homogeneous or a $\,\delta\,$-shell
field, and then extended to any rotationally invariant field
$\,B(x)\,$ which is of a compact support and does not change sign
\cite{CC}.

Our first aim here is to show that the last condition can be
substantially weakened and the rotational--invariance requirement can
be dropped altogether. We consider the standard two--dimensional Pauli
electron Hamiltonian \cite{Th},
\begin{equation} \label{Pauli}
H_P^{(\pm)}(A)\,=\,\left( -i\nabla-A(x)\right)^2\,\pm\,{g\over 2}
B(x) \,=\, D^*D\,\pm\,{1\over 2}(g\pm 2) B(x)\,,
\end{equation}
in natural units, $\,2m= \hbar= c= e= 1\,$; here $\,D:=
(p_1\!-A_1)+ i(p_2\!-A_2)\,$ and the two signs correspond to two
possible spin orientations. We are free to choose the magnetic flux
direction; if it points conventionally up we will be concerned
primarily with the operator $\,H_P^{(-)}(A)$ which describes electron
with the spin antiparallel to the flux. The magnetic field
$\,B= \partial_1A_2\!-\partial_2A_1\,$ is supposed to be integrable
and of a compact support $\,\Sigma\,$, with
\begin{equation} \label{flux}
F\,:=\, {1\over 2\pi}\, \int_{\Sigma} B(x)\, d^2x \,=\, N+\epsilon\,,
\end{equation}
where $\,\epsilon\in(0,1]\,$ and $\,N\,$ is a non--negative integer.
The quantity $\,F\,$, positive by assumption, is the total flux
measured in the natural units $\,(2\pi)^{-1}$.

Recall further that by the theorem of Aharonov and Casher
\cite{AC,Th} the operator $\,H_P^{(-)}(A)$ without an anomalous
moment, $\,g=2\,$, has in this situation $\,N\,$ zero energy
eigenvalues. The corresponding eigenfunctions are given explicitly by
\begin{equation} \label{ef's}
\chi_j(x)\,=\, e^{-\phi(x)} (x_1\!+\!ix_2)^j, \quad j=0,1,\dots,
N\!-\!1\,,
\end{equation}
where
\begin{equation} \label{phi}
\phi(x)\,:=\, {1\over 2\pi}\, \int_{\Sigma} B(y)\,\ln|x\!-\!y|\, d^2y
\,.
\end{equation}
Moreover, $\,\chi_N\,$ also solves the equation
$\,H_P^{(-)}(A)\chi=0\,$ representing a zero--energy resonance;
this follows from the fact that $\,\chi_j(x)= o(|x|^{-F+j})\,$ as
$\,|x|\to\infty\,$ --- \cf \cite[Sec.7.2]{Th}. \vspace{3mm}

\noindent
{\bf Theorem 1.} Under the stated assumptions, the operator
$\,H_P^{(-)}(A)\,$ has for $\,g>2\,$ at least $\,n_B\,$ negative
eigenvalues, where $\,n_B\,$ is the number of $\,j=0,1,\dots,N\,$
such that
\begin{equation} \label{condition}
\int_{\Sigma}B(x)\, e^{-2\phi(x)}\, r^{2j}\, d^2x\,\ge 0\,,
\end{equation}
where $\,r:=(x_1^2\!+\!x_2^2)^{1/2}$. In particular, there are at
least $\,n_B=N\!+\!1\,$ bound states if $\,B(x)\ge 0\,$.
\vspace{2mm}

\noindent
{\em Sketch of the proof:} It is based on a variational argument.
We employ the above mentioned zero--energy solutions to construct
a family of trial functions $\,\psi\,$ which make the quadratic form
$$
(\psi,H_P^{(-)}(A)\psi)\,=\, \int_{\R^2} |D\psi|^2 d^2x
\,-\,{1\over 2}(g\!-\!2) \int_{\R^2} B |\psi|^2 d^2x
$$
negative. Specifically, we choose
\begin{equation} \label{psi}
\psi_j(x)\,:=\, f_R(x)\chi_j(x) +\varepsilon h(x)\,,
\end{equation}
where $\,h\in C_0^{\infty}(\Sigma)\,$ and $\,f_R(x)= f\left(r\over R
\right)\,$ for a suitable function $\,f: \R_+\to \R\,$ such that
$\,f(u)=1\,$ for $\,u\le 1\,$. It is then straightforward to compute
the value of the energy form,
\begin{eqnarray*}
(\psi_j,H_P^{(-)}(A)\psi_j) &\!=\!& {1\over R^2}\, \int_{\R^2}
\left|f'\left(r\over R \right)\chi_j(x)\right|^2 d^2x \,+\,
\varepsilon^2\, \int_{\Sigma} |(Dh)(x)|^2 d^2x \\ \\
&\!-\!& {1\over 2}(g\!-\!2) \bigg\lbrace \int_{\Sigma}
B(x)|\chi_j(x)|^2 d^2x \\ \\ &\!+\!& 2\varepsilon \re\, \int_{\Sigma}
\bar h(x)B(x)\chi_j(x)\, d^2x
\,+\, \varepsilon^2 \int_{\Sigma} B(x)|h(x)|^2 d^2x \bigg\rbrace\,,
\end{eqnarray*}
where we have employed $\,D\chi_j=0\,$ and the fact that $\,h\,$ and
$\,f'\left(\cdot\over R \right)\,$ have disjoint supports. As we have
said, $\,\chi_j\in L^2\,$ for $\,j=0,\dots,N\!-\!1\,$. In this case
we put $\,f=1\,$ so the first term at the \rhs is absent. If $\,
\int_{\Sigma} B|\chi_j|^2 d^2x > 0\,$ we may set also
$\,\varepsilon=0\,$ to get a negative value. If $\,B\,$ is
non--negative, in particular, we obtain in this way
$\,(\psi_j,H_P^{(-)}(A)\psi_j)<0\,$ for $\,j=0,\dots,N\!-\!1\,$.

For a sign--changing $\,B\,$ the last integral might not be
positive. If it is zero, a bound state still exists: it is always
possible to choose $\,h\,$ in such a way that $\,\re\int_{\Sigma}
\bar h B\chi_j\, d^2x\ne 0\,$. For small $\,\varepsilon\,$ the
linear term prevails over the quadratic ones and the form can be
made negative by choosing properly the sign of $\,\varepsilon\,$.
Finally, for $\,j=N\,$ the Aharonov--Casher solution has to be
modified at large distances to produce a square integrable trial
function. We choose, \eg, $\,f\in C_0^{\infty}(\R_+)\,$ such that
$\,f(u)=0\,$ for $\,u\ge 2\,$. Using $\,|\chi_N(x)| =
o(r^{-\epsilon})\,$ we estimate the first term at the \rhs as
$$
{1\over R^2}\, \int_{\R^2} \left|f'\left(r\over R
\right)\chi_j(x)\right|^2 d^2x\,\le\, C\|f'\|^2_{\infty}
R^{-2\epsilon}
$$
for a positive $\,C\,$. If (\ref{condition}) is valid, one can achieve
in the same way as above that the sum of the other terms is negative;
it is then sufficient to set $\,R\,$ large enough to make the whole
\rhs negative.

We have thus constructed $\,n_B\,$ trial functions with the
desired property. They are linearly independent, since the same is
true for $\,\chi_j\,$ and the latter coincides with $\,\psi_j\,$
in $\,\BB_R\setminus\Sigma\,$. Consequently, the $\,\psi_j$'s for
which the requirement (\ref{condition}) is satisfied span an
$\,n_B\,$--dimensional subspace in $\,L^2(\R^2)\,$. \quad $\Box$
\vspace{2mm}

While the sufficient condition of Theorem~1 improves earlier
results, it is still too restrictive. We postpone discussing how
to optimize it to a subsequent paper. \vspace{2mm}

The situation becomes more complicated when the total flux is zero.
Here we will restrict ourselves to the particular case with a rotational
symmetry; then (\ref{Pauli}) can be replaced by a family of partial
wave Hamiltonians
\begin{equation} \label{partial wave Hamiltonian}
H_{\ell}^{(\pm)}=\,-\,{d^2\over dr^2} \,-\,{1\over r}\, {d\over dr}
\,+\, V_{\ell}^{(\pm)}(r)\,, \quad V_{\ell}^{(\pm)}(r):=\,
\left( A(r)+\, {\ell\over r} \right)^2 \pm\, {1\over 2}\,g B(r)
\end{equation}
on $\,L^2(\R^+,r\,dr)\,$. The angular component $\,A(r)\,$ of the
vector potential is now related to the magnetic field by $\,B(r)=
A'(r)+ r^{-1}A(r)\,$.

A typical situation with a vanishing flux arises when the field is
generated by an electric current vortex in the plane. The physical
appeal of such a problem stems in part from the fact that local
current vortices are common in transport of charged particles
\cite{ESF}. In the following we shall discuss this example. We
assume that the current is anticlockwise, $\,J(x)= \lambda J(r)
e_{\varphi}\,$.  Here $\,r,\varphi\,$ are the polar coordinates,
the total current is $\,\lambda \int_0^{\infty} J(r)\,dr\,$, and
the positive parameter $\,\lambda\,$ is introduced to control the
vortex ``strength".

It is necessary in this case to relax the compact--support
requirement on the magnetic field. We suppose that $\,J\,$ is
$\,C^2$ smooth and non--negative, $\,J(r)\ge 0\,$, and has the
following asymptotic behaviour,
\begin{equation} \label{asympt}
J(r)= ar^2+\OO(r^3)\quad {\rm and} \quad J(r)= \OO(r^{-3-\epsilon})
\end{equation}
for some $\,\epsilon>0\,$ at the origin and at large distances,
respectively. The corresponding vector potential is easily evaluated
\cite{Ja},
\begin{equation} \label{A}
A(r)\,=\, 4\lambda\, \int_0^{\infty} J(r')\, {r'\over r_<}\:
\left\lbrack K\left(r_<^2\over r_>^2 \right) -
E\left(r_<^2\over r_>^2\right)\right\rbrack\,dr'\,,
\end{equation}
where $\,K,\,E\,$ are the full elliptic integrals of the first and
the second kind, respectively, and the usual shorthands, $\,r_<:=
\min(r,r')\,$ and $\,r_>:= \max(r,r')\,$ are employed. In view of the
regularity of $\,J\,$ the integral is finite for every $\,r\,$,
because $\,E(\zeta)\,$ is regular at $\,\zeta=1\,$ and $\,K(\zeta)\,$
has a logarithmic singularity there.

Let us label the Pauli Hamiltonian (\ref{Pauli}) with the vector
potential (\ref{A}) and its partial--wave components (\ref{partial
wave Hamiltonian}) by the current strength $\,\lambda\,$.
\vspace{3mm}

\noindent {\bf Theorem 2.} Under the stated assumptions,
$\,\sigma(H_{\ell}^{(\pm)}(\lambda))= [0,\infty)\,$ for $\,\ell\ne
0\,$, while both $\,H_0^{(\pm)}(\lambda)\,$ exhibit a bound state
if $\,\lambda\,$ is small enough. On the other hand, each operator
$\,H_{\ell}^{(-)}(\lambda)\,$ has a negative eigenvalue for a
sufficiently large $\,\lambda\,$. \vspace{3mm}

\noindent
{\em Sketch of the proof:} By the regularity of $\,J\,$, the
effective potentials (\ref{partial wave Hamiltonian}) are $\,C^1$
smooth and
\begin{equation} \label{decay}
V_{\ell}^{(\pm)}(r)\,=\, {\ell^2\over r^2}\,+\, \lambda m\,
{2\ell\mp g\over 2r^3}\,+\, \OO(r^{-3-\epsilon})\,,
\end{equation}
as $\,r\to\infty\,$, where $\,m:= \pi\, \int_0^{\infty} J(r')\, r'^2
dr'\,$ is the dipole moment of the current for $\,\lambda=1\,$.
Consequently, the essential spectrum is not affected by the magnetic
field. We rewrite the potentials into the form
\begin{equation} \label{lambda dependence}
V_{\ell}^{(\pm)}(r)\,=\, \left( \lambda A_1(r)+\, {\ell\over r}
\right)^2 \pm\, {\lambda\over 2}\,g B_1(r)\,,
\end{equation}
where the indexed magnetic field refers to the value
$\,\lambda=1\,$. Since $\,H_{\ell}^{(\pm)}(\lambda)\,$ is nothing
else than the $s$--wave part of the two--dimensional Schr\"odinger
operator with the centrally symmetric potential (\ref{lambda
dependence}), it is sufficient to find eigenvalues of the latter.
If $\,\ell\ne 0\,$, the first term in (\ref{lambda dependence}) is
below bounded by $\,\lambda h(r)\,$ for a suitably chosen positive
function $\,h\,$ of compact support. Since the second term does
not contribute to $\,\int_0^{\infty} V_{\ell}^{(\pm)}(r)\,r\,dr\,$
which determines the weak--coupling behaviour, the result follows
from the standard condition \cite{Si} and the minimax principle.

While the above integral is positive in the case $\,\ell=0\,$ as
well for any $\,\lambda\ne 0\,$, this fact itself need not prevent
binding. A more careful Birman--Schwinger analysis up to the
second order in $\,\lambda\,$ is required: it shows that a weakly
coupled bound state exists if
\begin{equation} \label{weak bound}
\int_{\R^2} A(x)^2\, d^2x \,+\, {g^2\over 8\pi}\, \int_{\R^2\times
\R^2} B(x) \ln |x\!-\!x'|\, B(x')\, d^2x\, d^2x'\, < 0\,.
\end{equation}
Evaluating the last integral, we find that the condition is
satisfied for $\,g>2\,$. This rectifies an incorrect claim made in
\cite{BEZ}; a more detailed discussion on that point will be
presented in a forthcoming publication. The asymptotic behaviour
of the bound state energy (in the sense of \cite{Si}) is
\begin{equation} \label{weak asympt}
\epsilon(\lambda) \,\approx\, -\, \exp\left\{ -\left(
{\lambda^2\over 8} (g^2\!-4)\, \int_{\R^2} A(r)^2\,r\, dr
\right)^{-1} \right\}
\end{equation}
for both spin orientations (since $\,g\ne 2\,$, the second theorem
of \cite{AC} does not apply and the degeneracy may be lifted in
the next order).

On the other hand, the existence of the ``antiparallel" bound
states in a strong vortex follows from the behaviour of the
effective potential around the origin. We have
\begin{equation} \label{A small}
A(r)\,=\, \lambda\mu r + \alpha_0(r)\,, \qquad \mu\,:=\, \int_0^{\infty}
J(r')\, {dr'\over r'}\;;
\end{equation}
using (\ref{A}) and properties of the elliptic integrals we find
$\,\alpha_0(r)= \OO(r^2)\,$. This further implies
\begin{equation} \label{B small}
B(r)\,=\, 2\lambda\mu + \beta_0(r)\,, \qquad \beta_0(r)\,:=\,
\alpha_0'(r)+\, {1\over r} \alpha_0(r)\,=\,\OO(r)\,.
\end{equation}
Consider the case $\,\ell=0\,$. We substitute to (\ref{partial
wave Hamiltonian}) from (\ref{A small},\ref{B small}) and employ
the rescaled variable $\,u:= r\sqrt{\lambda}\,$. In this way
$\,H_0^{(\ell)}\,$ is unitarily equivalent to the operator $\,\lambda
A_{\lambda}\,$, where $\,A_{\lambda}= A_0+ W_{\lambda}\,$ on
$\,L^2(\R_+,u\,du)\,$ with
\begin{equation} \label{A_0}
A_0\,:=\,-\,{d^2\over du^2} \,-\,{1\over u}\, {d\over du}
\,- g\mu + \mu^2 u^2
\end{equation}
and
\begin{equation} \label{W}
W_{\lambda}(u)\,:=\,2\sqrt{\lambda} \mu\, u\alpha_0\left(u\over
\sqrt{\lambda} \right)\,+\, \lambda \, \alpha_0^2\left(u\over
\sqrt{\lambda} \right)\,-\, {1\over 2}\, g\, \beta_0\left(u\over
\sqrt{\lambda} \right)\,.
\end{equation}
The limit $\,\lambda\to\infty\,$ changes the spectrum substantially;
we have $\,\sigma_{ess}(A_{\lambda})=
\sigma_{ess}(\lambda A_{\lambda})= [0,\infty)\,$ for any $\,\lambda>0\,$,
while $\,A_0\,$ as the $s$--wave part of the two--dimensional harmonic
oscillator has a purely discrete spectrum. Nevertheless, one can
justify the use of the asymptotic perturbation theory for stable
(\ie, negative) eigenvalues of $\,A_0\,$; the fact that
$\,W_{\lambda}\to 0\,$ pointwise together with the resolvent identity
imply $\,A_{\lambda}\to A_0\,$ in the strong resolvent sense as
$\,\lambda\to\infty\,$ \cite{BEZ}. In that case there is a family of
$\,\nu_n(\lambda)\in \sigma(A_{\lambda})\,$ to any $\,\nu_n\in
\sigma_p(A_0)\,$  such that $\,\nu_n(\lambda)\to\nu_n\,$ \cite{Ka}.
The spectrum of $\,A_0\,$ is given explicitly by
\begin{equation} \label{HO spectrum}
\nu_n\,=\, \mu\left(4n+2-g \right)\,, \qquad n\,=\, 0,1,\dots\,,
\end{equation}
so $\,\nu_0\,$ is stable for $\,g>2\,$ and $\,A_{\lambda}\,$ has a
negative eigenvalue for $\,\lambda\,$ large enough. The analogous
argument applies to the case $\,\ell\ne 0\,$ , where the potential
in (\ref{A_0}) is replaced by $\,\mu^2 u^2 +\ell^2 r^{-2}
+\mu(2\ell\!-\!g)\,$, and one looks for negative eigenvalues among
$\,\nu_{n,\ell}= \mu \left(4n+ 2(|\ell|\!+\!\ell)+2-g\right)\,$.
The critical $\,\lambda\,$ at which the eigenvalue emerges from the
continuum is naturally $\,\ell$--dependent. $\quad\Box$
\vspace{2mm}



V.Z. thanks for the hospitality extended to him at NPI. The
research has been partially supported by GACR under the contract
202/96/0218.


\begin{thebibliography}{article}
%
\bibitem[AC]{AC}
Y.~Aharonov, A.~Casher: Ground state of a spin--1/2 charged particle in a
two--dimensional magnetic field, {\em Phys. Rev.} {\bf A19} (1979),
2641--2642.
\vspace{-1.8ex}
%
\bibitem[BEZ]{BEZ}
F.~Bentosela, P.~Exner, V.A.~Zagrebnov: Electron trapping by
a current vortex, {\em J. Phys.} {\bf A31} (1998), L305--311.
\vspace{-1.8ex}
%
\bibitem[CC]{CC}
R.M.~Cavalcanti, C.A.A.~de Carvalho: Bound states in spin-1/2 charged
particle in a magnetic flux tube, {\em J. Phys.} {\bf A31} (1998),
7061-7063.
\vspace{-1.8ex}
%
\bibitem[CFC]{CFC}
R.M.~Cavalcanti, E.S.~Fraga, C.A.A.~de Carvalho: Electron localization by a
magnetic vortex, {\em Phys. Rev.} {\bf B56} (1997), 9243--9246.
\vspace{-1.8ex}
%
\bibitem[E\v{S}SF]{ESF}
P.~Exner, P.~\v{S}eba, A.F.~Sadreev, P.~St\v{r}eda, P.~Feher:
Strength of topologically induced magnetic moments in a quantum
device, {\em Phys. Rev. Lett.}  {\bf 80} (1998), 1710--1713.
\vspace{-1.8ex}
%
\bibitem[Ja]{Ja}
J.D.~Jackson: {\em Classical Electrodynamics}, John Wiley, New York 1962.
\vspace{-1.8ex}
%
\bibitem[Ka]{Ka}
T.~Kato: {\em Perturbation Theory for Linear Operators}, Springer,
Heidelberg 1966.
\vspace{-1.8ex}
%
\bibitem[Mo]{Mo}
A.~Moroz: Single--particle density of states, bound states, phase--shift
flip, and a resonance in the presence of an Aharonov--Bohm potential, {\em
Phys. Rev.} {\bf A53} (1996), 669--694.
\vspace{-1.8ex}
%
\bibitem[Si]{Si}
B.~Simon: The bound state of weakly coupled Schr\"odinger operators in
one and two dimensions, {\em Ann.Phys.} {\bf 97} (1976), 279--288.
\vspace{-1.8ex}
%
\bibitem[Th]{Th}
B.~Thaller: {\em The Dirac equation}, Springer, Berlin 1992.
\vspace{-1.8ex}
%
   \end{thebibliography}
\end{document}